\documentclass[twocolumn,preprintnumbers,superscriptaddress,longbibliography]{revtex4-2}
\usepackage[dvipsnames]{xcolor}
\definecolor{persianblue}{rgb}{0.11, 0.22, 0.73}
\definecolor{mediumpersianblue}{rgb}{0.0, 0.4, 0.65}
\definecolor{persianred}{rgb}{0.8, 0.2, 0.2}
\definecolor{persianplum}{rgb}{0.44, 0.11, 0.11}
\usepackage[normalem]{ulem}
\usepackage[colorlinks=true,citecolor=persianblue,linkcolor=persianblue, urlcolor=persianblue]{hyperref}
\usepackage{amsmath}
\usepackage{amssymb}
\usepackage{soul}
\usepackage{fontawesome}
\usepackage{graphicx}
\usepackage{enumitem}
\usepackage{hyperref}
\usepackage{physics}
\usepackage[dvipsnames]{xcolor}
\usepackage{soul}
\usepackage{xspace,slashed}
\usepackage{qcircuit}
\usepackage[OT1]{fontenc}
\usepackage{cmbright}
\DeclareFontShape{OT1}{cmss}{m}{it}{<->ssub*cmss/m/sl}{}

\usepackage{subfigure}
\usepackage{listings}

\lstdefinestyle{codestyle}{
    commentstyle=\color{Gray},
    keywordstyle=\bfseries\color{Magenta},
    stringstyle=\color{RedOrange},
    identifierstyle=\color{black},
    basicstyle=\ttfamily\footnotesize,
    breakatwhitespace=false,
    breaklines=true,
    captionpos=b,
    keepspaces=true,
    numbersep=5pt,
    showspaces=false,
    showstringspaces=false,
    showtabs=false,
    tabsize=2,
    frame=tb,
    framerule=0.1pt
}
\lstdefinelanguage{MyPython}[]{Python}{
  morekeywords={as},
}
\newcommand{\Qibolab}{\texttt{Qibolab}\xspace}
\newcommand{\Qibo}{\texttt{Qibo}\xspace}
\newcommand{\Qibocal}{\texttt{Qibocal}\xspace}

\newcommand{\RX}{$\text{R}_\text{X}$\xspace}
\newcommand{\RZ}{$\text{R}_\text{Z}$\xspace}

\definecolor{sergi}{rgb}{1,0,.5}

\definecolor{ingo}{rgb}{.8,.5,0}

\begin{document}

\title{Qibocal: an open-source framework for calibration of self-hosted quantum devices}

\preprint{TIF-UNIMI-2024-16, CERN-TH-2024-160}

\newcommand{\MIaff}{TIF Lab, Dipartimento di Fisica, Universit\`a degli Studi di
  Milano, Italy.}

\newcommand{\INFNUNIMI}{INFN, Sezione di Milano, I-20133 Milan, Italy.}

\newcommand{\UNIMIB}{Dipartimento di Fisica, Universit\`a di Milano-Bicocca, 3 20126 Milano, Italy.}

\newcommand{\INFNUNIMIB}{INFN - Sezione di Milano Bicocca, 3 20126 Milano, Italy.}

\newcommand{\BQT}{Bicocca Quantum Technologies (BiQuTe) Center, 3 20126, Milano, Italy.}

\newcommand{\TII}{Quantum Research Center, Technology Innovation Institute, Abu Dhabi, UAE.}

\newcommand{\CERNaff}{CERN, Theoretical Physics Department, CH-1211
  Geneva 23, Switzerland.}

\newcommand{\genericCERNaff}{European Organization for Nuclear Research (CERN), Geneva 1211, Switzerland.}

\newcommand{\UB}{Departament de F\'isica Qu\`antica i Astrof\'isica and Institut de Ci\`encies del Cosmos (ICCUB), Universitat de Barcelona, Barcelona, Spain.}

\newcommand{\CQT}{Centre for Quantum Technologies, National University of Singapore, Singapore.}

\newcommand{\NTU}{Division of Physics and Applied Physics, School of Physical and Mathematical Sciences, Nanyang Technological University, 21 Nanyang Link, Singapore 637371, Singapore.}

\newcommand{\SNR}{Medical Physics, IRCCS San Raffaele Scientific Institute, Milan, Italy.}

\newcommand{\BRA}{Instituto de F\'isica de S\~ao Carlos, Universidade de S\~ao Paulo -- S\~ao Carlos (SP), Brasil.}

\newcommand{\JKU}{Department of Quantum Information and Computation at Kepler (QUICK), Johannes Kepler Universität Linz, 4040 Linz, Austria.}

\author{Andrea Pasquale}
\thanks{These authors contributed equally to this work.}
\affiliation{\MIaff}
\affiliation{\TII}
\affiliation{\INFNUNIMI}

\author{Edoardo Pedicillo}
\thanks{These authors contributed equally to this work.}
\affiliation{\MIaff}
\affiliation{\TII}

\author{Juan Cereijo }
\affiliation{\TII}

\author{Sergi Ramos-Calderer}
\affiliation{\TII}
\affiliation{\UB}

\author{Alessandro Candido}
\affiliation{\CERNaff}

\author{Gabriele Palazzo}
\affiliation{\TII}
\affiliation{\SNR}

\author{Rodolfo Carobene}
\affiliation{\UNIMIB}
\affiliation{\INFNUNIMIB}
\affiliation{\BQT}

\author{Marco Gobbo}
\affiliation{\UNIMIB}
\affiliation{\INFNUNIMIB}
\affiliation{\BQT}

\author{Stavros Efthymiou}
\affiliation{\TII}

\author{Yuanzheng Paul Tan}
\affiliation{\NTU}

\author{Ingo Roth}
\affiliation{\TII}

\author{Matteo Robbiati}
\affiliation{\MIaff}
\affiliation{\genericCERNaff}

\author{Jadwiga Wilkens}
\affiliation{\TII}
\affiliation{\JKU}

\author{Alvaro Orgaz-Fuertes}
\affiliation{\TII}

\author{David Fuentes-Ruiz}
\affiliation{\TII}

\author{Andrea Giachero}
\affiliation{\UNIMIB}
\affiliation{\INFNUNIMIB}
\affiliation{\BQT}

\author{Frederico Brito}
\affiliation{\BRA}
\affiliation{\TII}

\author{Jos\'e I. Latorre}
\affiliation{\TII}
\affiliation{\CQT}
\affiliation{\UB}

\author{Stefano Carrazza}
\affiliation{\CERNaff}
\affiliation{\MIaff}
\affiliation{\INFNUNIMI}
\affiliation{\TII}

\begin{abstract}
    Calibration of quantum devices is fundamental to successfully
    deploy quantum algorithms on current available quantum hardware.
    We present \Qibocal, an open-source software library to perform calibration and characterization of
    superconducting quantum devices within the \Qibo framework. \Qibocal completes
    the \Qibo middleware framework by providing all necessary tools to easily
    (re)calibrate self-hosted quantum platforms.
    After presenting the layout and the features of the library, we give an overview on some of the protocols
    implemented to perform single and two-qubit gates calibration.
    Finally, we present applications involving recalibration and monitoring of superconducting platforms.
\end{abstract}

\maketitle

% \tableofcontents

\section{Introduction}

Recent developments in quantum technologies have demonstrated
promising applications of quantum algorithms on hardware~\cite{utility}.
However, a major challenge preventing larger applications is the noise
affecting current quantum systems, which reduces their fidelity.
In fact, the error correction mechanisms~\cite{error_correction, surface_code} required for the advent of
fault-tolerant devices demand an error threshold far lower than what can currently be
achieved by large-scale quantum systems.

To achieve and maintain current state-of-the-art fidelities, characterization and calibration software
has become as crucial as well-designed and fabricated quantum hardware.
Furthermore, maintaining accurate calibration of such devices requires a significant daily time investment.
The noise affecting current devices results in shifts~\cite{shifts}
in optimal parameter configurations, which must be addressed and corrected through
appropriate recalibration experiments~\cite{kelly2018physical}.

\begin{figure}[ht]
  \includegraphics[width=0.5\textwidth]{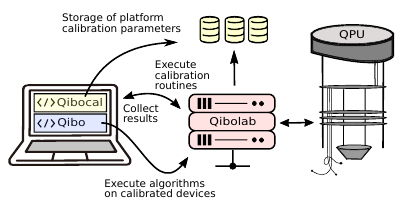}
  \caption{\label{fig:scheme}Schematic representation of \Qibocal's role in the hybrid-quantum
  operating system proposed within the \Qibo framework.}
\end{figure}

Current cloud providers of quantum hardware, including \texttt{IBM Quantum}~\cite{qiskit2024},
provide limited access to the code necessary for deploying quantum algorithms, lacking
details about the software responsible for calibration.

Recently, several open-source libraries dedicated to quantum device calibration and characterization have
emerged~\cite{forestbenchmarking, qiskit-experiments, Nielsen_2020, scqubits1, qubex, pulser, tergite}.
However, these tools are typically tied to specific control electronics, hardware platforms, architectures,
or proprietary technologies, and therefore do not support the seamless translation of experiments across
different instruments.

To the best of our knowledge, no currently available library provides the full set of capabilities required
to calibrate a quantum device through a device-agnostic backend. Existing tools instead address only subsets
of this problem: some focus primarily on benchmarking routines available only on proprietary
platforms~\cite{qiskit-experiments, tergite}; others target highly specialized hardware
architectures~\cite{qubex, pulser}; some are restricted to gate-based characterization
protocols~\cite{forestbenchmarking}; others are limited to simulation~\cite{scqubits1}; and some address only
a specific layer of the characterization stack~\cite{Nielsen_2020}.

\Qibocal, by contrast, is designed to accommodate heterogeneous control electronics while providing the functionality
needed to maintain an up-to-date calibration of a self-hosted quantum device.

To make software for controlling self-hosted devices widely accessible,
we recently introduced \texttt{Qibolab}~\cite{qibolab_paper, pedicillo2024opensourceframeworkquantumhardware} as
a submodule of \texttt{Qibo}~\cite{qibo_paper, Efthymiou_2022, Qibo_proceeding, pasquale2024statevectorsimulationqibo}, an open-source middleware framework for quantum computing.
This module includes primitives for managing the experimental setups required for quantum computing.
With \texttt{Qibolab}, both low level experiments and \texttt{Qibo} circuits can be executed
on self-hosted quantum devices, which are increasingly becoming available for in-house use in research institutions.

Control over instruments alone is insufficient for successfully deploying quantum algorithms on hardware.
Proper calibration and characterization of the entire system must be performed.
For superconducting qubits, this involves performing a series of experiments
designed to optimize the microwave pulses that implement native gates and measurements.
\texttt{Qibocal}, which is based on both \texttt{Qibo} and \texttt{Qibolab},
was developed to ease the deployment of such calibration protocols on superconducting devices.

The development of \Qibocal was originally motivated by several practical
limitations encountered in self-hosted calibration software for superconducting
quantum platforms. In many experimental setups, calibration routines are
implemented as ad-hoc scripts that tightly couple data acquisition, analysis,
and parameter updates. As the number of available characterization and
calibration experiments grows, these approaches tend to produce large and
difficult-to-maintain codebases, making it challenging to reuse experiments,
track dependencies between routines, and integrate new calibration protocols.
Establishing a robust software infrastructure is therefore a crucial step
toward building reliable and scalable calibration tools.

Despite significant progress in the automation of quantum hardware operation,
the calibration of quantum processors still relies heavily on human
intervention. In this setting, \Qibocal aims to streamline the experimental workflow
 of data streaming, experiment management, and parameter updates. In
addition, \Qibocal provides a flexible interface for composing calibration
protocols, enabling experimentalists to construct custom calibration workflows
tailored to specific devices or experimental objectives. While
it is possible to define recalibration schemes using a direct acyclic graph
(DAG)~\cite{kelly2018physical}, specific applications can benefit from a more
comprehensive dependency scheme including, for example, optimization loops.

In Fig.~\ref{fig:scheme}, we show schematically the role of \Qibocal within the \Qibo framework.

Overall, we want to emphasize the lack of an all encompassing calibration solution designed for self-hosted quantum devices. This problem is what \Qibocal tackles, and sets the stage for a calibration framework that is able to agnostically compare, and be deployed on, quantum devices regardless of accoutrements. To properly present this framework, 
the paper is organized as follows. In Sect.~\ref{sect:technical} we present a detailed overview
of the \texttt{Qibocal} library as of version \texttt{0.1.0}. Sect.~\ref{sect:calibration} lists some of the
characterization and calibration protocols available in the library.
In Sect.~\ref{sect:applications} we showcase the capabilities of the library
by presenting applications using \texttt{Qibo}.
Finally, in Sect.~\ref{sect:outlook} we draw our conclusions and describe future developments.

\section{Technical specification}
\label{sect:technical}

\Qibocal is the software layer within the \Qibo~\cite{qibo_paper} framework responsible for the
characterization and calibration of quantum processors controlled through \Qibolab~\cite{qibolab_paper}.
At its core, \Qibocal provides a collection of experiments, referred to as \emph{protocols}, designed to
extract physical parameters of quantum devices and to fine-tune calibration settings in order to optimize
operations. Beyond protocol execution, the library offers tools to manager experiments,
store and share results, and automatically update the quantum processing unit (QPU) configuration according
to newly calibrated parameters.

\begin{figure}
  \includegraphics[width=0.9\columnwidth]{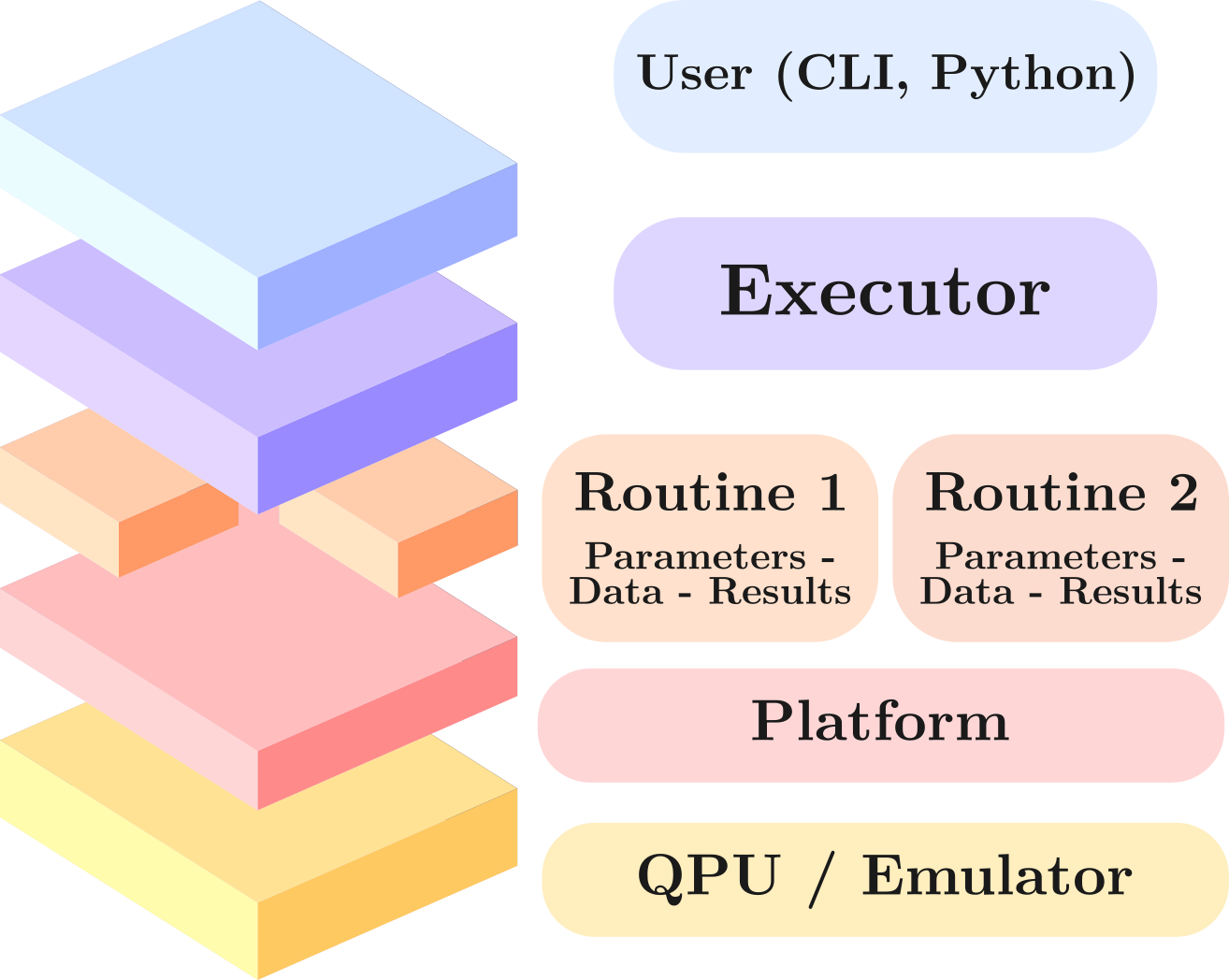}
  \caption{Overview of the \Qibocal workflow. Users interact via a suitable interface. The Executor manages multiple Routines, each containing Parameters, Data, and Results. The Platform is updated by the Executor after each Routine and deployed to the QPU or emulator.}
  \label{fig:layer_scheme}
\end{figure}

This section provides an overview of the package structure and a typical \Qibocal workflow,
illustrated in Fig.~\ref{fig:layer_scheme}.

Users can interact with \Qibocal either through Python scripts or via the command line interface (CLI).
The workflow is managed by an \texttt{Executor} object, which can handle multiple experiments, represented as \texttt{Routine}
objects. Each experiment contains substructures (\texttt{Parameters}, \texttt{Data}, and \texttt{Results}), which are used as
interface between the various stages of executions, representing the data flow. This segmentation in precise stages, with
specific intermediate entities, is also beneficial to preserve and reproduce results, through serialization of the
intermediate objects. Which in turn is also valuable for detailed inspection of successful and failing workflows.

The \texttt{Routine} is deployed directly to the QPU (or to the \Qibolab emulator), while the \texttt{Executor} updates the
calibration configuration accordingly.

Detailed descriptions of the structure, workflows, and interfaces are provided in the following subsections.

Most calibration routines in \Qibocal rely on model-based analysis of experimental data.
Typical examples include sinusoidal fits for Rabi oscillations and exponential fits for relaxation
and coherence measurements. These fitting
procedures enable the automatic extraction of optimal calibration parameters from experimental
datasets, providing a lightweight and interpretable data-driven calibration pipeline.
Machine learning techniques are currently employed mainly for readout classification in the IQ plane,
where different models can be trained and evaluated according to several performance metrics.

At the same time, the framework does not impose a rigid separation between acquisition and analysis.
Protocols can be extended to implement more complex acquisition strategies in which the experiment
is dynamically adapted based on intermediate results. In this sense, the standard pattern consisting
of a fixed acquisition stage followed by a fitting procedure should be regarded as a common usage
rather than a strict requirement. This topic is tackled with more details in Sec.~\ref{sec:api}.

The modular structure of \Qibocal also makes it suitable for integration with more advanced
optimization strategies. Since calibration protocols expose both the acquisition layer and the
resulting figures of merit, they can be naturally embedded within adaptive optimization loops by
an external orchestrator~\cite{qiboml}.

\subsection{Software design}

\begin{figure*}[ht]
  \includegraphics[width=0.7\textwidth]{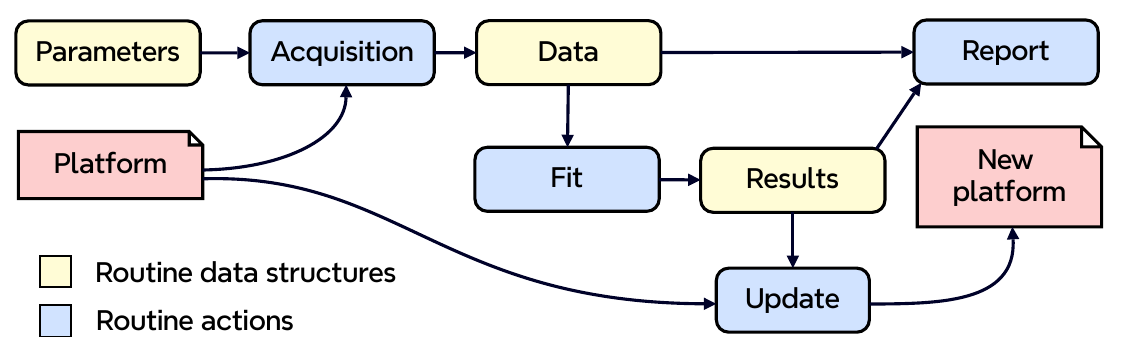}
  \caption{Dependence scheme for a \Qibocal \texttt{Routine}.}
  \label{fig:routine}
\end{figure*}

We can generally split an experiment in three different stages:
\begin{itemize}
\item \textbf{Data acquisition:} execution of measurements on the quantum device, ranging from low-level
pulse sequences or circuit runs to more complex experimental procedures.
\item \textbf{Data processing:} classical computation performed after measurements, when access to
quantum hardware is no longer required.
\item \textbf{Data visualization:} generation of plots and tables summarizing results, typically compiled
into an automatically generated report.
\end{itemize}

These stages and their relations are illustrated in Fig.~\ref{fig:routine}.

\begin{table}
\caption{Core classes involved in a \texttt{Routine}, their roles within the execution workflow. In the third column,
an example of what every object would contain in a standard Rabi experiment.}
\label{tab:routine_classes}
\centering
\begin{tabular}{p{1.8cm} p{2.2cm} p{4cm}}
\hline
\textbf{Class} & \textbf{Role} & \textbf{Example:\newline Rabi oscillations} \\
\hline
\texttt{Parameters} & Experiment configuration & Pulse amplitude sweep range, fixed pulse duration, number of shots, relaxation delays \\
\texttt{Data} & Raw measurements container & Measured probabilities to measure te qubit in the excited state \\
\texttt{Results} & Processed quantities & Estimated $\pi$-pulse amplitude \\
\hline
\end{tabular}
\end{table}

Table~\ref{tab:routine_classes} summarizes the main classes involved in the definition of a \texttt{Routine} object and their
roles within the execution workflow.

The external input of a \texttt{Routine} is provided through a \texttt{Parameters} object, which contains
all configuration values required during acquisition. These may include, for example, the frequency sweep
range of a spectroscopy experiment, the number of repetitions, or relaxation delays between measurements.

Raw measurements produced during acquisition are stored in a \texttt{Data} object. By default, datasets are
saved in binary format for efficient storage and loading (primarily using NumPy-compatible formats), while
auxiliary metadata are serialized as \texttt{JSON} files. Nevertheless, \Qibocal allows users to define
custom serialization backends, provided that corresponding loading and dumping methods are supplied.

Information extracted from \texttt{Data} during post-processing is collected into a \texttt{Results} object and serialized alongside the data.
This object contains a compact representation of the relevant physical quantities inferred from the
experiment and is subsequently used to update the QPU configuration with newly calibrated parameters.

Implementing a new protocol therefore typically requires defining the corresponding \texttt{Parameters},
\texttt{Data}, and \texttt{Results} classes together with acquisition and processing logic. Reusability of the
classes definitions among similar experiments is possible and encouraged. Detailed
guidelines and examples are provided in the official documentation~\cite{docs}.

\subsubsection{Command line execution}

\begin{figure*}
  \includegraphics[width=1\textwidth]{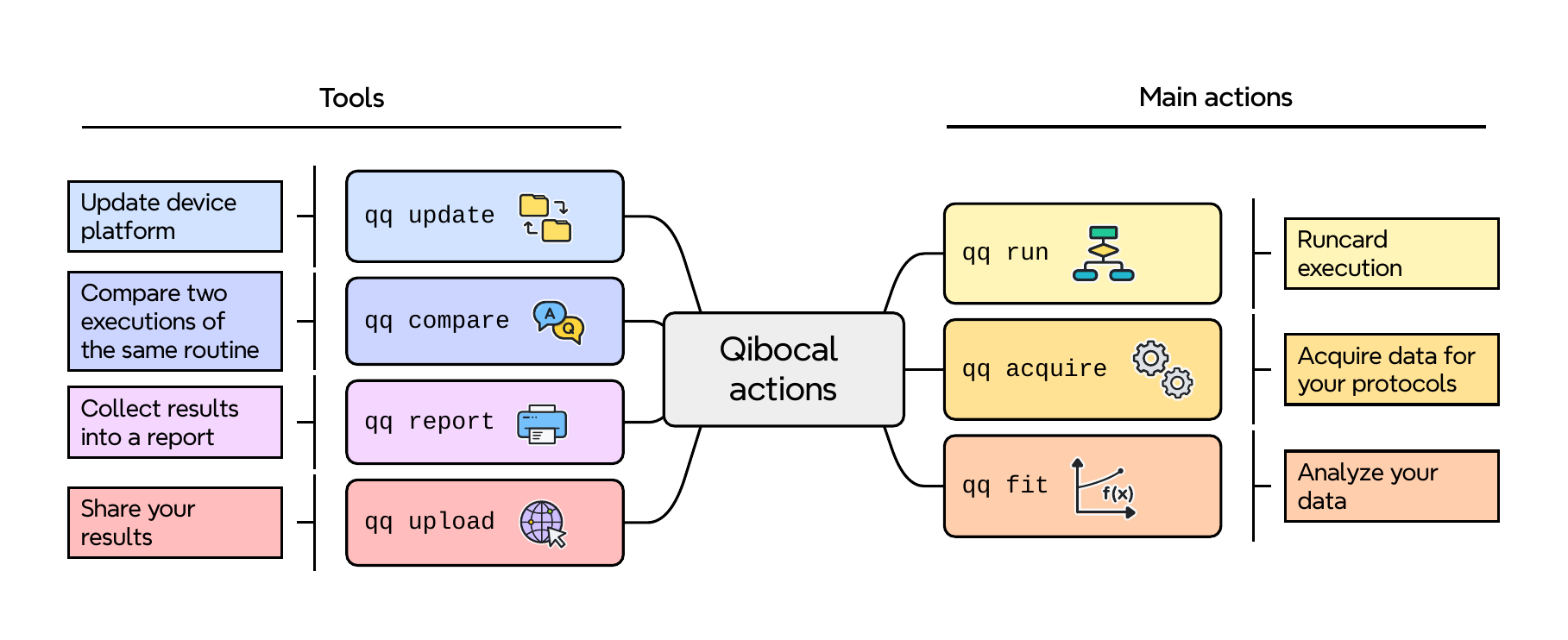}
  \caption{\label{fig:cli}Qibocal CLI workflow.}
\end{figure*}

\Qibocal provides a streamlined way to launch protocols through a command line interface (CLI),
as summarized in Fig.~\ref{fig:cli}.
All protocol-related information prior to execution—mainly the \texttt{Parameters} and QPU
configurations—is serialized into an \emph{experiment runcard}.

The \texttt{Runcard} is then deserialized by an \texttt{Executor} object, which handles:
\begin{enumerate}
    \item Instantiation of the corresponding \texttt{Routine}
    \item Acquisition and post-processing of the data
    \item Update of the QPU configuration with calibrated parameters
\end{enumerate}

Through the CLI, users can select which steps to execute, restricting the workflow to the necessary functionality at a given moment.
Data acquisition, processing, and reporting can be run independently via \texttt{qq acquire}, \texttt{qq fit}, and \texttt{qq report}, respectively. The additional command \texttt{qq update} adds the possibility to import experiment results to be used for future experiments.
A full mode that combines all steps is also available through \texttt{qq run}, intended as the general-purpose command for complete execution.

An \emph{experiment runcard} may contain multiple experiments, which are executed sequentially.
In this case, QPU calibration parameters are updated after each protocol, so subsequent experiments operate with a configuration reflecting the parameters calibrated by the previous experiment.

This sequential approach is particularly suitable for monitoring applications and short recalibration programs.
For more complex, non-sequential workflows, it is recommended to use the Python API directly (see Sect.~\ref{sec:api}).

\subsubsection{A calibration program}
\label{sec:api}

\begin{figure*}
  \includegraphics[width=0.7\textwidth]{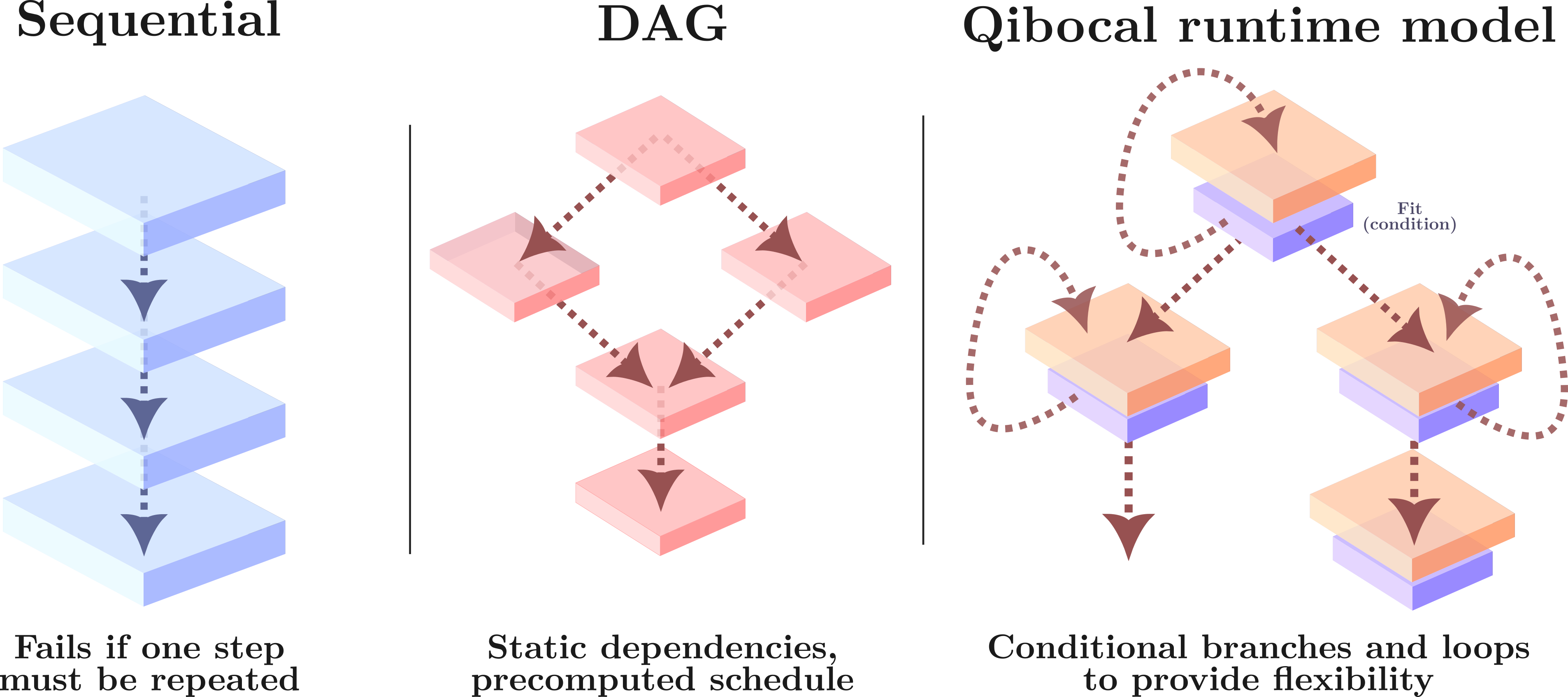}
  \caption{\label{fig:dag_comparison}Comparison of execution models for calibration workflows: sequential pipelines, static DAG scheduling, and dynamic programmatic execution. Only the latter supports runtime decisions, conditional branches, and adaptive repetitions.}
\end{figure*}

To perform a complete calibration, sequential execution is not flexible enough, since it
does not fully account for failed fits or experimental quirks that may appear in any
experiment.
In the management of complex workflows, this flexibility would be beneficial to address
the variety of possible situations. Especially in the development of general and
portable solutions.

To support this use case, graphs have often been used~\cite{kelly2018physical}. In
particular, directed acyclic graphs (DAG) have been considered an optimal device to
represent a general calibration process.
What makes DAGs convenient is that they straightforwardly map to a given execution, as
they define dependency relations.
These relations could be resolved in a clear and possibly optimized way, e.g.\
exploiting its topological sorting.

Not all types of execution are easily represented by a DAG.
The prerequisite for that is to plan the whole execution process ahead of time.
Indeed, being directed and without cycles, they can only proceed forward, defining a clear timeline for the
execution.
This contrasts with typical procedural programming model: the execution flow is not
necessarily linear, but it can jump between function calls, conditional, and loops.
E.g., DAGs purposefully prevents cyclic dependencies~\cite{kelly2018physical},
approximating them with their unwrapping in layers of precision, but also any other
runtime conditional (which lies at the foundation of a loop).

The graph approach is then well-suited for some scenarios, but not necessarily an
all-weather solution.
Two example workflows that may be compared are the model-driven calibration process,
based on specific low-dimensional protocols, predicting the optimal value of few control
parameters out of grid scans' fits, and some data-driven optimization, which could
perform a search of optimal parameters in a large-dimensional space.
The first is the typical calibration process, which is often simple to describe, and
quite portable across different setups. In this case, graphs are a suitable solution.
However, once this is applied, and the system is already in the vicinity of the optimal
operating point, iterative techniques may be experimented to further improve the
calibration. And by their nature of being iterative, they are hard to represent as
\textit{acyclic} graphs.

Therefore, \Qibocal is designed to provide this flexibility and naturalness for the
execution process, embedding the execution specification in the most common high-level
representation: a programming language.
This is achieved by mapping the atomic operation, i.e.\ the protocol, to a callable
function, the computation primitive of a procedural programming language, which
accept parameters as input and return the post-processing results as the method's
output. Then, they can be composed together, as represented in
Fig.~\ref{fig:dag_comparison}.

Beyond making the operations flexible and the specification process simpler, leveraging
the programming languages' expressivity, it enhances the integration process with other
tools, as the result of many protocols can be processed together with the aid of
external libraries in the middle of the calibration workflow, and even directly integrate
with further tools (monitoring, data provider, etc.).

Protocols are then limited to physical experiments, defining a specific control
sequence.

The exact executor and protocol API is still evolving, but a typical \Qibocal script
would look like:

\vspace{0.2cm}
\begin{lstlisting}[language=MyPython]
  def recurse(executor, par):
    output = executor.protocol_inner(..., par=par)
    if condition(output):
        return output
    return recurse(executor, output.some_par)

  executor = Executor(...)
  executor.connect()

  output_a = executor.protocol_a(...)
  output_rec = recurse(executor, ...)
  ...

  executor.disconnect()
  executor.save()
\end{lstlisting}

For example, this versatile interface simplifies the implementation of some
early iterative optimization experiments, using a fidelity (e.g. the
randomized-benchmarking one) as a target figure of merit, in combination with common
general-purpose optimizer, such as those provided by the SciPy
package~\cite{2020SciPy-NMeth}.
An alternative workflow involved iterating over a fixed sequence of simple protocols
(e.g. Rabi, Ramsey, single-shot readout classification), studying the convergence
properties of the iterative execution.

The programmatic interface, while implicitly suggesting certain workflow patterns, is
deliberately designed to remain agnostic to specific scaling solutions for large-device
calibration. Consequently, it does not provide a pre-defined approach for this task.
Given that scaling calibration constitutes a distinct challenge—best addressed
independently of experiment design and execution—the primary objective is to ensure a
clear separation of concerns between these two domains.

\subsection{Reports}

\Qibocal provides a set of tools to facilitate the generation, visualization, and sharing
of experiment results.

At the end of a program executed via the CLI, \Qibocal can automatically produce a web page
summarizing all protocols that were run. The content of this report is fully customizable
through the \texttt{report} function, which is part of the \texttt{Routine} interface
illustrated in Fig.~\ref{fig:routine}. Each protocol is expected to generate a
collection of plots and tables, which are then aggregated into the final report.
In the \texttt{Routine} definition it is also possible to provide \texttt{HTML} strings
in order to fully customize the output for specific experiments.

Reports generated by \Qibocal can be easily shared by uploading them to a dedicated server
using the CLI command \texttt{qq upload}. While a full database solution would be the
better approach for larger-scale deployments, the current method is sufficientt for most use
cases. Future versions may expose this functionality directly within custom protocol scripts.

\begin{figure}
  \includegraphics[width=\columnwidth]{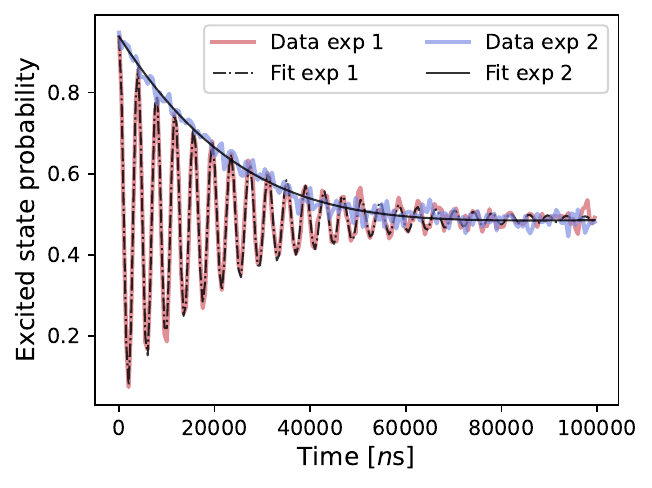}
  \caption{Comparison between two Ramsey experiments executed using the command \texttt{qq compare} in \Qibocal. In blue we present a first run of the experiment.
  The presence of oscillations is an indication of the fact that the frequency of the drive pulse is not aligned with the qubit frequency. In red
  we run the same experiment after correcting the qubit frequency.}
  \label{fig:qq-compare}
\end{figure}

Additionally, \Qibocal allows graphical comparison of two reports using the CLI command
\texttt{qq compare}. This produces a combined \texttt{HTML} report in which plots and
tables from both runs are juxtaposed, enabling rapid assessment of changes in the
experimental setup. An example of this comparison functionality is shown in
Fig.~\ref{fig:qq-compare}.

\subsection{Hardware support}

As mentioned in the introduction, \Qibocal is designed as a high-level orchestration
framework for quantum device calibration, built atop Qibolab --- a hardware abstraction
layer that standardizes interactions with diverse quantum control hardware. By
leveraging Qibolab, Qibocal inherently supports all hardware backends integrated within
its framework, including FPGA-based systems and real-time feedback capabilities. This
architecture enables Qibocal to focus on calibration workflows while delegating
low-level hardware interactions to Qibolab, ensuring compatibility with both local and
cloud-based setups. Notably, real-time operations (e.g., pulse sequencing) are handled
during the acquisition phase, while post-experiment analysis (e.g., fitting) occurs
offline. This separation streamlines integration with academic and industrial
infrastructures, without requiring Qibocal itself to implement hardware-specific logic.

\section{Calibration protocols}
\label{sect:calibration}

\begin{figure*}[ht]
  \includegraphics[height=0.25\textheight]{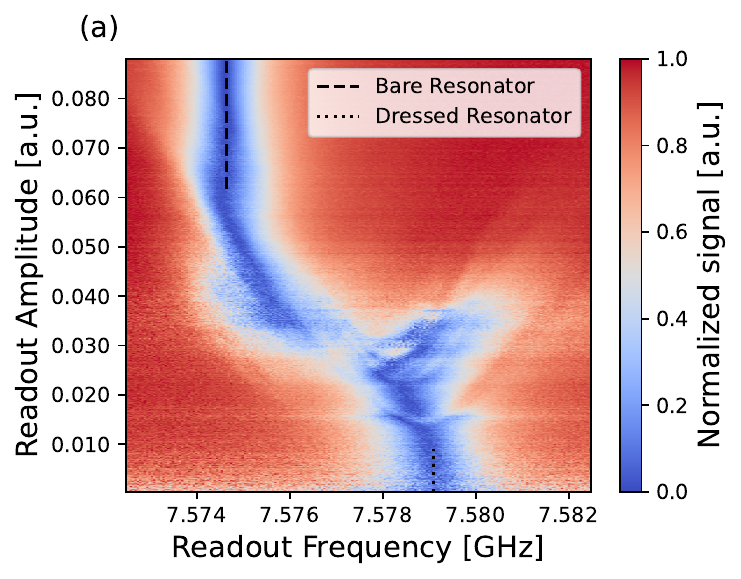}
  \includegraphics[height=0.25\textheight]{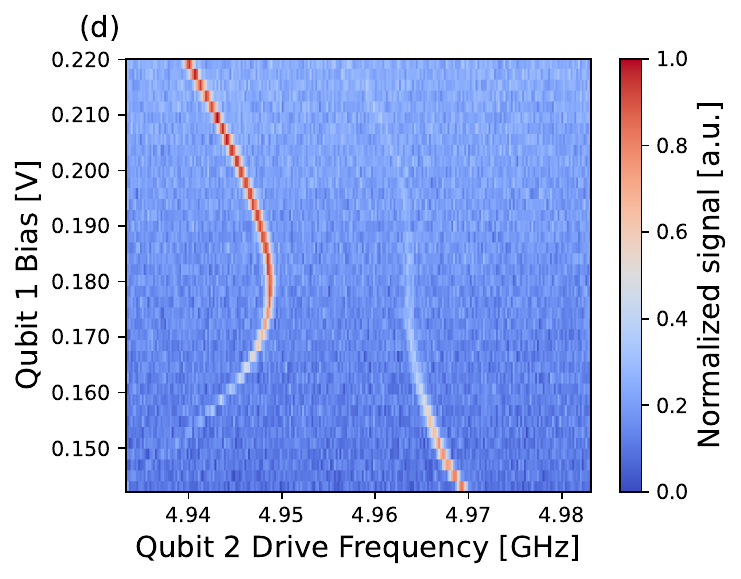}
  \includegraphics[height=0.25\textheight]{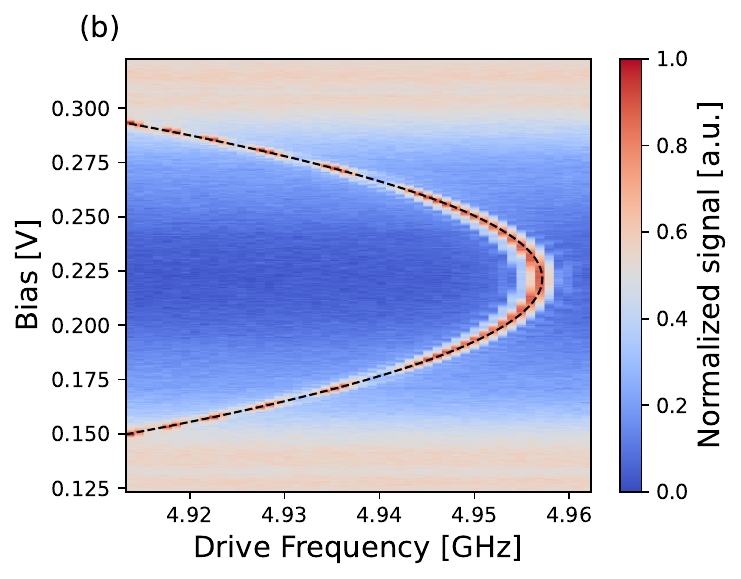}
  \includegraphics[height=0.25\textheight]{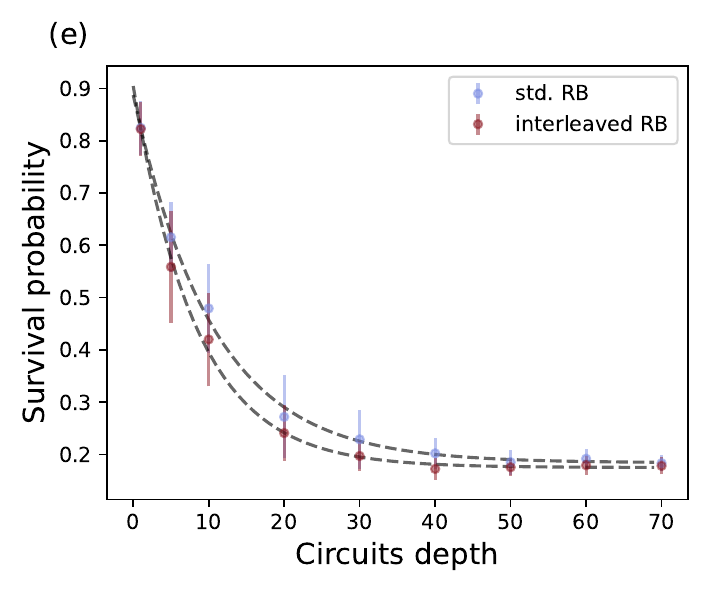}
  \includegraphics[height=0.25\textheight]{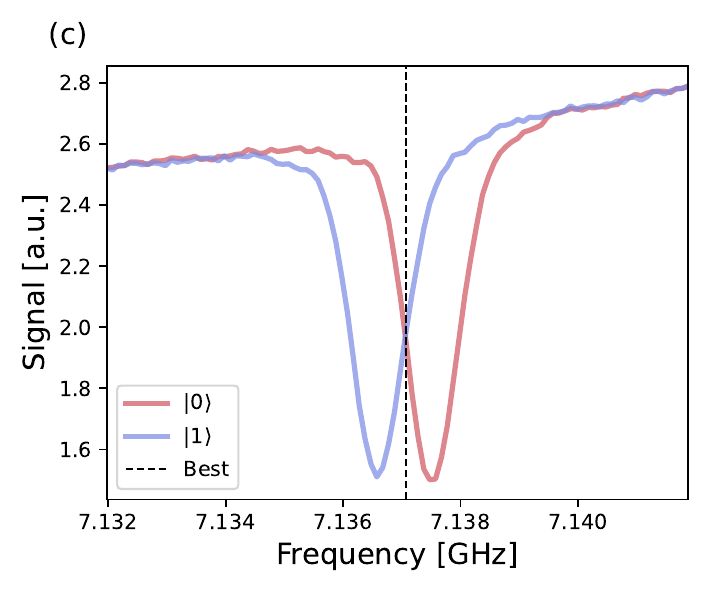}
  \includegraphics[height=0.25\textheight]{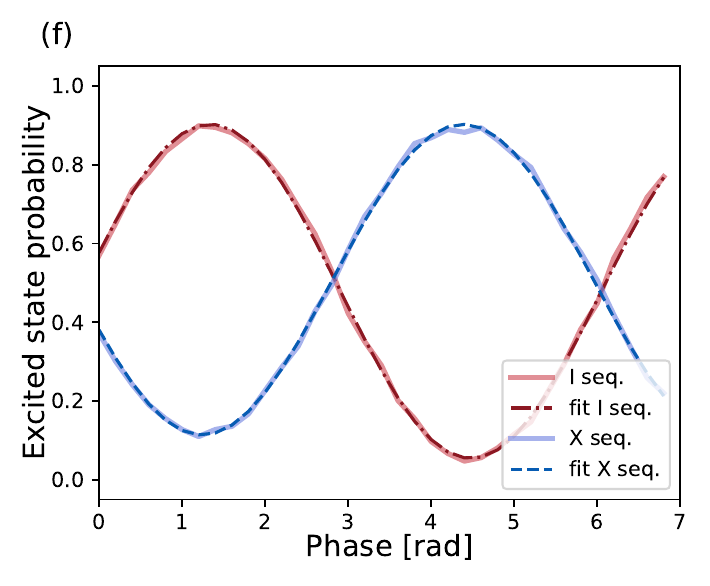}
  \caption{Gallery of six \Qibocal protocols. \emph{(a)} Measurement of bare and dressed resonator frequency.
  \emph{(b)} Qubit frequency measured as a function of the external bias line.  The dashed line shows the expected
  frequency.
  \emph{(c)} Transmission coefficient for different readout frequencies for a qubit
  prepared in state $\ket{0}$ (red) or $\ket{1}$ (blue). The dashed black line corresponds to the readout frequency which maximizes
  the separation between the two states.
  \emph{(d)} Avoided crossing measured between two qubits.
  \emph{(e)} Standard and interleaved randomize benchmarks with two qubits gate.
  \emph{(f)} Correction of dynamical single-qubit phases after the application of a flux pulse implementing a
  CZ gate~\cite{Rol_2019}.}
  \label{fig:prot}
\end{figure*}

The code base introduced in the previous section can be harnessed to deploy calibration protocols
developed using \Qibolab primitives. In this library, we include an extensive suite of calibration protocols,
not only as useful tools for the calibration of superconducting qubit devices, but also as a starting point
for developing new, personalized routines.

In this section we highlight some of the \Qibocal features, describing a few of the
available routines. To this purpose, we introduce them by following a standard, basic,
calibration procedure~\cite{GaoGuide}.

\subsection{Single qubit calibration}

The first step towards fully controlling a qubit is the calibration of single qubit gates.
More precisely, single qubit \RX gates together with \RZ rotations are sufficient to access
any single qubit rotation~\cite{Krantz_2019}.
Although \RZ rotations can be performed in several ways~\cite{Manenti:2023zzn},
the most effective method consists in a virtual implementation~\cite{McKay_2017}
based on shifting the phase of subsequent pulses during circuit compilation.
For this reason, \RX rotations and measurement gates are the targets
for single qubit calibration.

To assess the resonator and qubit frequencies of an unknown device,
\Qibocal provides ad hoc protocols for single and two-tone spectroscopies.
For resonators of a 2D notch variety we provide fits~\cite{10.1063/1.4907935} not to
only retrieve parameters like the frequency of the resonator, but also quality factors and impedance mismatches.
A Lorentzian fit is provided for other resonator types and for quick qubit recalibration schemes, where we expect
to probe the resonator in a narrow frequency range.
The frequency of the qubit can be extracted with a two-tone spectroscopy by fitting the transmitted signal with a Lorentzian fit.
Moreover, the protocol controls the amplitude of the tone driving the qubit, allowing the same routine to probe higher energy transitions
such as two photon transitions between the ground state and the second excited state.

Next is the calibration of a coherent $\pi$-pulse, a complete rotation from state $\ket{0}$ to $\ket{1}$.
This task requires the calibration of the amplitude and the duration
of a microwave pulse, and it is usually achieved through Rabi experiments~\cite{Krantz_2019},
several versions of which can be found in \Qibocal.
Standard experiments to extract parameters related to the relaxation
time $T_1$ and coherence times $T_2^*$ and $T_2^{\rm Echo}$ are also provided.
\Qibocal also offers protocols to fine-tune drive pulse parameters, such as Ramsey experiments for correcting
the frequency of the drive pulse or sequences aimed at amplifying the error on the amplitude of the drive pulse,
which we refer to as \emph{flipping}~\cite{flipping}.

Thanks to \Qibolab, these protocols, and in general most calibration routines in \Qibocal,
support the acquisition of not only the integrated and demodulated readout signal,
i.e., for each shot we extract the in-phase and quadrature components (IQ-plane),
but also single-shot readout, from which we can easily retrieve probabilities.

This second mode needs to be calibrated first, and \Qibocal offers several tools to help
maximize single-shot readout fidelity. Standard single-shot classification is available, aimed at providing the
parameters required for automatic discrimination within typical control electronics. Furthermore, \Qibocal is able
to train different machine learning models to classify states in the IQ-plane and offers different metrics
to choose from~\cite{pedicillo2023benchmarkingmachinelearningmodels}.

After running all protocols listed above, the user should be able to have
a first calibration of single qubit gates, however, better results can be achieved by running more complex experiments.
As the literature suggests~\cite{PhysRevLett.103.110501}, and \Qibocal includes,
calibration of the DRAG pulse will modify the envelope of \RX pulse rotations reducing leakage,
and finding the readout frequency that maximizes the distance of the signals generated by
the ground and the excited state in the IQ plane.

\subsection{Two qubit gates calibration}

\Qibocal includes the necessary tools to extend the calibration to two-qubit gates.
Various calibration schemes have been explored in the literature~\cite{Manenti:2023zzn},
but currently \Qibocal supports the calibration of two qubit interactions based on flux tunable qubits~\cite{Koch_2007},
including chips with tunable couplers~\cite{PhysRevApplied.10.054062, PhysRevApplied.14.024070}.
The standard procedure consists in sending a flux pulse through a dedicated line to the qubit,
which shifts its resonant frequency close to the one of a neighboring qubit,
allowing for a swap or controlled-phase interaction depending on the initial
state preparation~\cite{Krantz_2019}.
The inclusion of couplers has been proven useful to reduce the ZZ coupling~\cite{couplers}. However, such architecture
requires dedicated experiments to calibrate the couplers to switch on and off the interaction between the qubits.

One of the main challenges in implementing two-qubit gates is the presence of
distortions in the flux pulses, which originate from the limited bandwidth of
the control electronics and transmission lines and from electronic components
like bias tees. These distortions not only modify the waveform experienced by
the qubit but can also introduce long-timescale effects, where the flux at the
qubit at a given time depends on previous flux excursions. A common strategy to
mitigate this problem is to pre-distort the pulses according to the measured
response of the system. This predistortion is implemented by filtering the
waveform produced by the AWG through a combination of finite impulse response
(FIR) and infinite impulse response (IIR) digital filters. The short-timescale
distortions are evaluated through the Cryoscope experiment~\cite{cryoscope},
while the long-timescale ones are evaluated through a spectroscopy
experiment~\cite{lpd}.

For flux tunable calibration, \Qibocal has specific routines to identify the
interaction points for the $\text{iSWAP}$ and $\text{CZ}$ gates.
Fig.~\ref{fig:prot}b shows an example of qubit flux spectroscopy;
after performing the fit we are able to extract the qubit-flux
dependency (for more details see Sec.~\ref{sec:t1_t2_ro_flux}) and find the
operational flux point. In fact, if we move
the flux point of the first qubit in order to be in resonance with the second one
we can observe the typical avoided crossing shown in Fig.~\ref{fig:prot}d. Once we have
found the two-qubit interaction point, we can execute a Chevron-like experiment.
By sweeping the flux pulse amplitude and the flux pulse duration, we can assess the
pulse parameters required for a controlled two-qubit interaction. Such protocols exhibit
a Chevron-like plot routine~\cite{Krantz_2019}.
This illustrates \Qibocal's ability to fit and recognize parameters within
two-dimensional data.

To further fine-tune the degree of interaction and calibrate a specific two-qubit
gate, \Qibocal provides additional routines to extract the correct
conditional rotation.
It also takes into account additional parameters such as remnant dynamical
single qubit phases, as in Fig.~\ref{fig:prot}f, and leakage to non-computational states~\cite{Rol_2019}.

While the previous experiments are common to all possible flux pulse waveforms,
\Qibocal also provides specific protocols for the calibration of the \texttt{SNZ} pulse~\cite{snz},
which reduces the leakage and the effect of long-timescale distortions,
improving the repeatability of the gate. The randomized benchmarks with \texttt{SNZ} pulse is shown in Fig.~\ref{fig:prot}e.

Regarding two-qubit gates with couplers, we have implemented specific
protocols to find the operational point of the couplers, i.e., where the
qubits' interaction is active.
To achieve the calibration, we sweep the amplitude of a flux pulse applied to the coupler and the duration of the qubit flux pulse
to tune the two-qubit gate's pulse sequences.

\subsection{Qubit benchmarking}

Within the \Qibocal library, complex benchmarking techniques are available to properly gauge the result of a
calibration suite. Including benchmarking capabilities in the workflow of \Qibocal is crucial,
as they provide the relevant figures of merit that the calibration should achieve.

Although most of these protocols are ultimately implemented through circuit execution
(e.g., using \Qibo), their integration within \Qibocal is motivated by the
broader calibration workflow. In this context, the availability of
interoperable building blocks with a unified interface allows the user to combine
benchmarking, characterization, and calibration routines within a single
framework.

Moreover, these benchmarking protocols can be interleaved within a calibration
suite to properly track the improvement of fitted parameters, and their output
can be used as cost functions within an optimization loop. In this setting,
\Qibo-based workflows are naturally embedded within \Qibocal protocols, where
the circuit execution constitutes only one layer of a larger infrastructure
designed to support systematic chip characterization and calibration.

Tomographies on different levels are also available as \Qibocal routines
following Ref.~\cite{DiCarlo_2009}. While useful outside the context of
calibration, they are an invaluable tool to certify the correctness of the
calibration parameters and extract more information from the resulting quantum
states and processes. More advanced characterization techniques, such as
gate-set tomography, could also be integrated within the same framework by
implementing the corresponding experimental sequences and post-processing
routines.

\begin{figure*}[ht]
  \includegraphics[height=0.188\textheight]{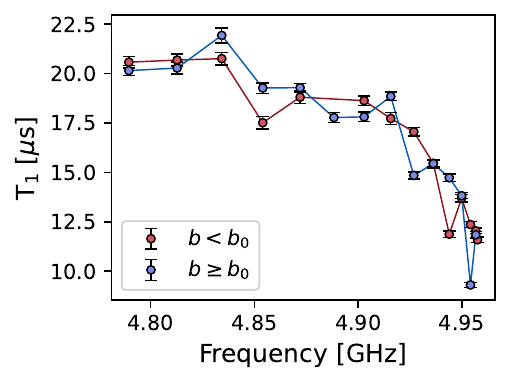}
  \includegraphics[height=0.188\textheight]{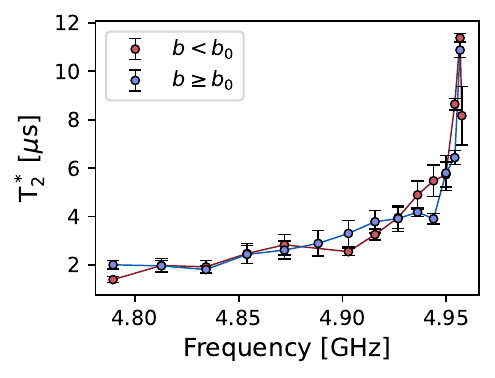}
  \includegraphics[height=0.188\textheight]{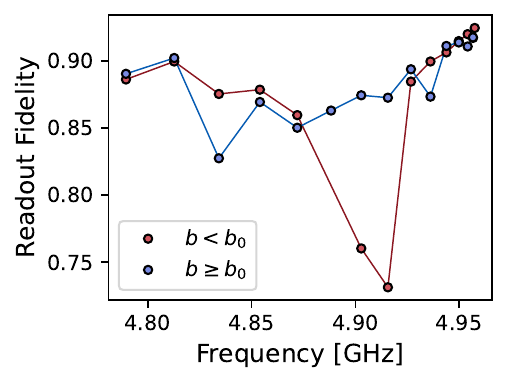}
  \caption{\label{fig:t1_flux}Measurements  of $\text{T}_1$,  $\text{T}_2^*$ and readout fidelity as the qubit is operated
  at different frequencies. The blue curve indicates that the detuning is induced by increasing the flux, while the red
  curve corresponds to the case where we are decreasing flux. For each point we follow the recalibration procedure described in Sect.~\ref{sec:t1_t2_ro_flux}.
  The behavior seen for T$_1$ measurements is consistent with the increase of the qubit's Purcell protection
  (see \cite{Schoelkopf2008} for a systematic study about the spontaneous emission of Transmon qubits).
   }
\end{figure*}

\subsection{Experiments}

The \Qibocal tools designed for calibration protocols can also be used on more complex experiments.
Moreover, the visualization and reporting that \Qibocal provides,
as well as its structured acquisition and fitting features can be readily adapted to fit any measurement.

\Qibocal also provides two experiments to quantify the entanglement of the system.
One experiment performs CHSH inequalities~\cite{PhysRevLett.23.880} over a range of
measurement settings to assert the entanglement generated by two-qubit gates in
the system.
The other experiment goes further, and performs three qubit Mermin
inequalities~\cite{10.1119/1.12594} to determine the ability of the system
to produce multi-partite entanglement.
Their purpose is twofold. They verify the calibrated parameters in a complete
end-to-end application, and they serve as models for the connection of
experiments through the \Qibocal pipeline.

Any measurement, regardless of its complexity, when coded within the \Qibo and
\Qibolab framework can be included in a calibration stack.
This way, by interconnecting calibration and experiments, \Qibocal can
automatically perform calibration protocols to improve the results of
the experiment immediately before said experiment is executed.

\subsection{Automatic recalibration}

This combination of protocols can be used as a tool for recalibration.
\Qibocal is able to feed the calibrated parameters from one routine to the next,
either sequentially or through non-trivial connections. The flexibility
offered through the calibration  program API presented in Sect.~\ref{sec:api} allows
for the creation of custom calibration schemes, including the possibility
to deviate from a default pipeline based on the outcome of the previous experiments.
Launching protocols directly through Python also enables the
possibility to code ad-hoc stopping conditions based on specific threshold values.
The verification tools available, such as tests designed to ensure the reliability
of the fitted parameters, can turn this process into an automatic recalibration
procedure.

\subsection{Running experiments on emulated platforms}
Thanks to the integration with \Qibolab, \Qibocal can be used to run experiments
on any \Qibolab-compatible platform. Among these a particular relevance is given to platforms
which emulate the behavior of a quantum processing unit, by solving the corresponding Master Equation of
the system which includes the time-dependent Hamiltonian generated by the control pulses.

In particular, starting from a noiseless Hamiltonian, it is possible to start launching experiments with
\Qibocal in an ideal environment before deploying them on a real quantum processing unit.
Being able to run experiments on an emulated platform brings several advantages, including the possibility
to test the calibration protocols and verify both the measurement and the post-processing steps.

Future work involving the emulator includes narrowing down parameter range for calibration routines performed
on the actual device. By running the calibration and emulating certain routines, the total time required to bring up
a new device can be cut down. As an example, through mismatching of the Rabi frequency and sampling rate of
the actual instrument, rotation errors of the control pulses will also manifest in the emulator.
This issue is naturally addressed by the \emph{flipping} protocol, but can also be performed in the emulator
as a first pass to find the approximate amplitude detuning.

Running on an emulation comes also with some drawbacks including the computational time required
to run the simulation particularly when targeting multi-qubit systems.
Currently the emulator has been tested with a system of split-transmons capacitively coupled, for testing most of
the experiments in \Qibocal it is usually enough to have a system with two qubits.

\section{Qibocal in action}
\label{sect:applications}

\subsection{Coherence at different bias points}
\label{sec:t1_t2_ro_flux}
As an example of how \texttt{Qibocal} can be used to write custom experiments
involving several protocols we compute the value of $T_1$, $T^*_2$ and readout
fidelity for a qubit biased at different flux points.
Although such experiment could be performed in a simpler way~\cite{klimov2018fluctuations} for the purpose of
showcasing the library capabilities we have recalibrated the qubit at each flux point.
Before the experiment, we characterize how the qubit frequency changes with the
flux by running a qubit flux spectroscopy
where we fit the data with the standard approximation for those qubits, see Eq.~1 in~\cite{Barrett_2023} for example.

Our calibration procedure for each flux point involves
the following steps:
\begin{enumerate}[noitemsep]
  \item update the qubit frequency according to the approximation used to fit the data;
  \item execute the Rabi experiment to recalibrate the drive pulse amplitude;
  \item execute the single-shot classification to run routines in single-shot readout acquisition mode;
  \item execute the Ramsey experiment to fine-tune the drive frequency;
  \item repeat the single-shot classification to fine-tune the readout;
  \item measure $T_1$, $T^*_2$ and readout fidelity.
\end{enumerate}
This experiment has been performed on a qubit built by QuantWare and controlled
using a Quantum Machines~\cite{quantum_machines} cluster through \texttt{Qibolab}.
The results are displayed in Fig.~\ref{fig:t1_flux}. We can observe that
the qubit exhibits a reasonably good $T_1$ value at the sweetspot
of around 10 $\mu$s; roughly at 150 MHz we can observe $T_1$ values reaching up to
20 $\mu$s.
$T^*_2$ peaks at the sweetspot and deteriorates as we increase the detuning, as expected.
Additionally, the readout fidelity reaches its maximum at the sweetspot.

\subsection{Pulse optimization with randomized benchmarking}
\begin{figure}[tb]
  \includegraphics[width=\linewidth]{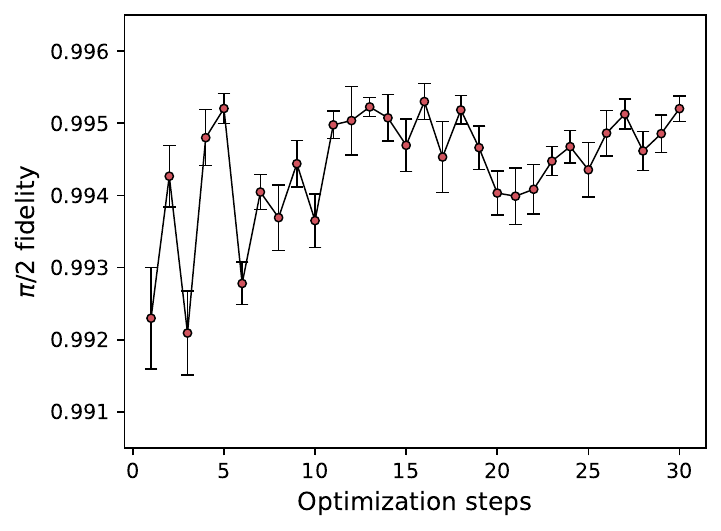}
  \caption{\label{fig:nelder_mead_rb}
  The $\tfrac\pi2$-pulse fidelity extracted from randomized benchmarking (RB) is used as the objective function of a Nelder-Mead optimization of the $\tfrac\pi2$-pulse amplitude and DRAG parameter.
  The resulting $\tfrac\pi2$-pulse fidelity after each step in the optimization is shown.
  The optimization results in an increase in fidelity with few evaluations of the cost function.
  }
\end{figure}

A common technique in the control of quantum devices is the optimization of pulse shapes based on high-level performance metrics.
A simple example of the more general approach of \textit{quantum optimal control}~\cite{Koch:OptimalControl:2022}, is
to search for pulse parameters maximizing the gate fidelities obtained from randomized benchmarking  experiments \cite{Helsen:Framework:2022} as the objective function \cite{Egger:OptimalControl:2014,Kelly:OptimalControl:2014}.
\texttt{Qibocal} is a flexible tool in the development and benchmarking of such techniques.
As a basic example, we implement a gradient-free Nelder-Mead optimization loop for the $\tfrac\pi2$-pulse
using \texttt{Qibocal}'s Clifford randomized benchmarking routine and the device parameter update mechanisms.
Fig.~\ref{fig:nelder_mead_rb} shows the improvement in the randomized benchmarking decay parameter with the number of Nelder-Mead steps proposing values for the pulse amplitude and DRAG parameter of the pulse.

\subsection{Re-calibration after changes in flux background}

Changes in the system and environment parameters, e.g.\ the flux background,
can require frequent re-calibration of super-conducting qubit.
Such workflows for such re-calibration protocol can be combined from \texttt{Qibocal} routines.
As example, we perform multiple re-calibrating of $\tfrac\pi2$ pulse using the following steps:
\begin{enumerate}[noitemsep]
  \item Ramsey spectroscopy for fine-tuning the drive frequency;
  \item single-shot classification for fine-tuning the readout;
  \item Rabi experiment to re-calibrate drive pulse amplitude;
  \item single-shot classification for fine-tuning the readout.
\end{enumerate}

\begin{figure}[tb]
  \includegraphics[width=1\linewidth]{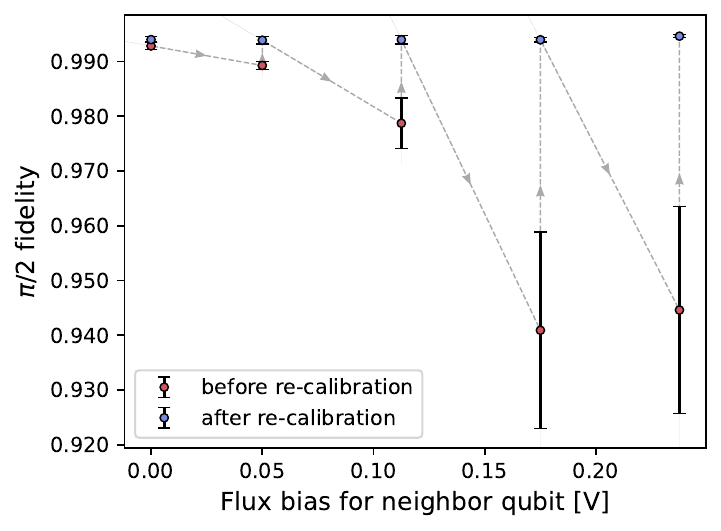}
  \caption{
    \label{fig:detune_and_recalibration_rb} Re-calibration of a qubit's
    $\tfrac\pi2$-pulse after (controlled) changes of its flux background.
    Applied voltage to flux line of neigbouring qubit and randomized benchmarking
    $\tfrac\pi2$-fidelity before (red) and after (blue) re-calibration against time.
    The gray dashed arrows describes how the calibration changes over time.
    Flux bias is relative to the qubit's calibrated sweetspot.
  }
\end{figure}

We use a standard Clifford randomized benchmarking experiment to monitor the $\tfrac\pi2$-pulse fidelity.
To simulate a controlled drift of the qubit frequency we change the flux background by sending a bias to the flux line of the connected qubit.
Fig.~\ref{fig:detune_and_recalibration_rb} shows a time line of the fidelities when applying different biases before and after re-calibration.

\begin{figure*}[ht]
  \includegraphics[width=\textwidth]{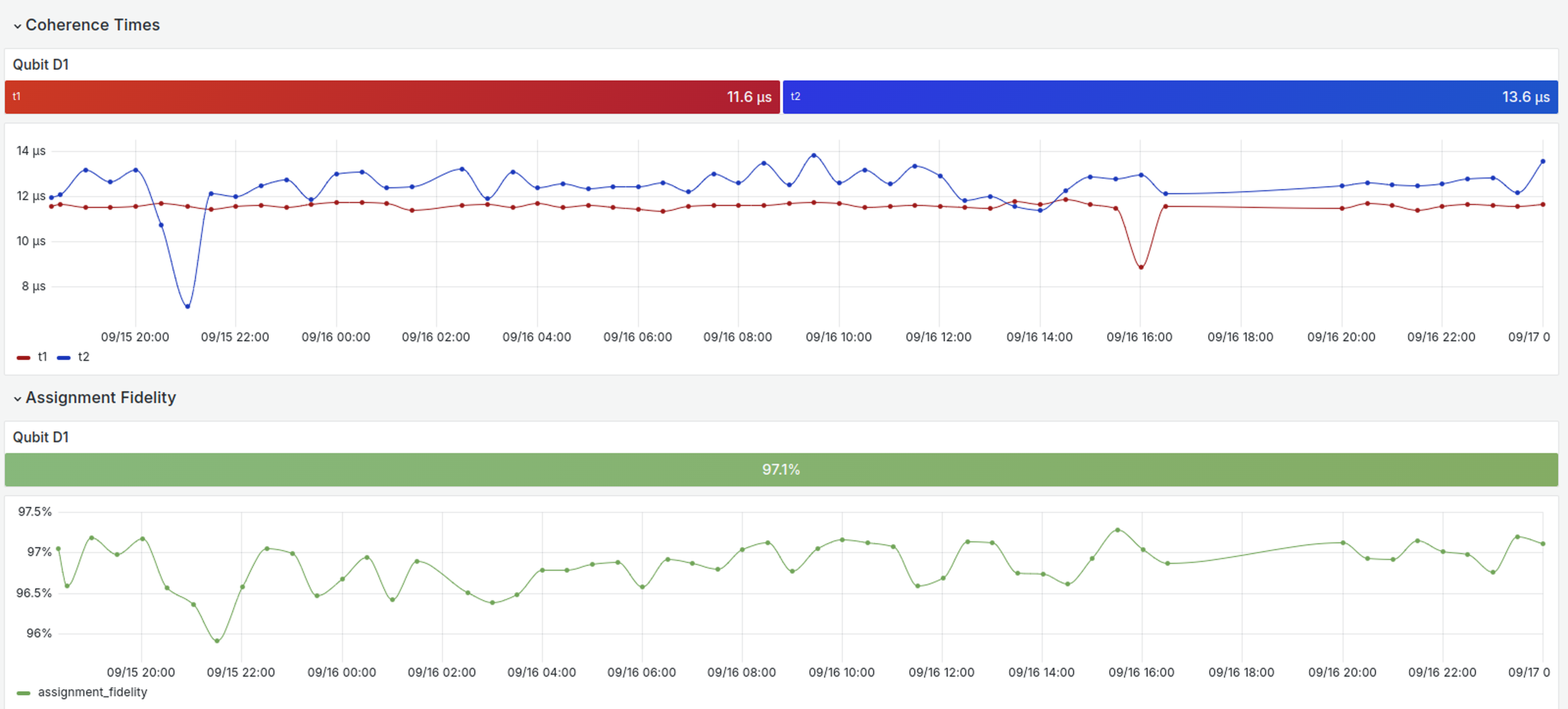}
  \caption{Monitoring of coherence times and readout fidelity using \texttt{Qibocal}.}
  \label{fig:grafana_screenshot}
\end{figure*}
\subsection{Monitoring qubit calibration}

\texttt{Qibocal} can be used in monitoring tools, such as Grafana~\cite{Grafana}.
In this specific case Docker containers were used, in order to ensure the easiest portability of such tools.
In particular, a Grafana Docker container was set up, automatically configuring the layout of the dashboard,
with the possibility to also install third-party plugins.
Monitoring containers can be deployed at regular time intervals, measuring several qubit metrics.
In the example of Fig.~\ref{fig:grafana_screenshot}, $T_1$, $T_2^*$ and readout fidelity from one qubit
were monitored every thirty minutes, with all measurements saved
in a dedicated container running PostgreSQL~\cite{postgresql}.
More complex workflows can be set up, including simultaneously monitoring multiple qubits and recalibrating
the chip if any metrics fall below a certain threshold.
Additionally, the modular structure of Docker containers allows for more containers,
to measure QPU usage or cryostat temperature.

\section{Outlook}
\label{sect:outlook}

In this paper we introduced \Qibocal, an open-source software library for
calibration and characterization of superconducting quantum devices. \Qibocal is
based on \Qibo, a full-stack software framework which provides a simple interface to
define circuit-based quantum algorithms via custom backends,
\textit{i.e.}~dedicated plugin software libraries which deploy algorithms on
specific hardware.

The release of \Qibocal increases the usability of \Qibo as a quantum middleware
framework by providing specialized tools to (re)calibrate self-hosted quantum
devices thanks to the seamless integration with \Qibolab platform and
instruments configuration.
\Qibocal aims to kickstart the software standardization of calibration
protocols for quantum devices by reducing code redundancy between research
groups and laboratories operating self-hosted quantum hardware platforms.

We described the software components and tools implemented in release
\texttt{0.1.0}, with a focus on the protocols required for the
calibration and characterization of single- and multi-qubit superconducting
devices.
Furthermore, we presented three examples of applications using \Qibocal that
demonstrated the utility of this library for quantum technology research:
coherence at different bias points, calibration stability against noise and
monitoring qubit calibration.

In future releases of \Qibocal, we intend to extend its capabilities by
defining custom and efficient calibration protocols for multi-qubit devices with
a larger number of qubits. Indeed, thanks to the modularity of the library, we
have the possibility to adapt and scale the API for large-scale systems.

Another promising direction is the integration of machine learning
and adaptive optimization techniques within the calibration pipeline,
enabling automated tuning strategies for increasingly large-scale
quantum processors.
Although in the current release of \Qibocal there are calibration experiments for two-qubit gates involving $\text{CZ}$ or $\text{iSWAP}$,
we are working to add new experiments to support also architectures with $\text{CNOT}$ as a native gate, implemented through
cross resonance~\cite{Paraoanu_2006}.
Even though our current work focuses on superconducting platforms, reflecting the present scope
of the \Qibo framework, the modular design of \Qibocal makes it extensible to
other quantum computing platforms, such as trapped ions, neutral atoms, or photonics. Extensions
to these platforms could be contributed by the community or by experimental collaborators as the
\Qibo control software evolves and new hardware becomes accessible.
We believe that with the release of \Qibocal, \Qibo becomes an even more unique
and useful tool for the quantum computing community, reducing the software
development effort for researchers in simulation, hardware control and
calibration.

The code implementing the \Qibocal module is available at:
\begin{center}
\small
  \href{https://github.com/qiboteam/qibocal}{\texttt{\faGithub\,\,https://github.com/qiboteam/qibocal}}.
\end{center}

All the data used in this manuscript are available at~\cite{data}:
\begin{center}
\small
  \href{https://github.com/qiboteam/qibocal-paper-data}{\texttt{\faFolderOpen\,\,https://github.com/qiboteam/qibocal-paper-data}}.
\end{center}

\acknowledgments This project is supported by TII's Quantum Research Center. The
authors thank all \Qibo contributors for helpful discussion. M.R.\,is supported
by CERN's Quantum Technology Initiative (QTI) through the Doctoral Student
Program. M.R.\,and S.C.\,thanks the CERN TH hospitality during the elaboration of
this manuscript.
J.W.\,acknowledges financial support from the Austrian Research Promotion Agency under Contract No. 897481 (High-Performance integrated Quantum Computing).
A.G.\,acknowledges support by the Horizon 2020 Marie Sklodowska-Curie action (H2020-MSCA-IF GA No.101027746).
R.C.\,and M.G.~acknowledge support by PNRR MUR projects PE0000023-NQSTI.

\newpage

\bibliography{references}

\end{document}